# Tuning Charge Density Wave in the Transition from Magnetically Frustrated Conductor to Ferrimagnetic Insulator in Carbon Nanowire within Boron Nitride Nanotube

[1]Chi Ho Wong*, [4]Zong Liang Guo, [1]King Cheong Lam, [1]Chun Pong Chau, [1]Wing Yu Chan, [3]Chak-yin Tang, [4]Yuen Hong Tsang, [2]Leung Yuk Frank Lam, [2]Xijun Hu

[1]Division of Science, Engineering and Health Studies, School of Professional Education and Executive Development, The Hong Kong Polytechnic University, Hong Kong, China

[2]Department of Chemical and Biological Engineering, The Hong Kong University of Science and Technology, Hong Kong, China

[3]Department of Industrial and Systems Engineering, The Hong Kong Polytechnic University, Hong Kong, China

[4]Department of Applied Physics, The Hong Kong Polytechnic University, Hong Kong, China

Email: roy.wong@cpce-polyu.edu.hk

**Abstract:**

The emergence of exotic charge density wave (CDW) alongside ferrimagnetism materials opens exciting new possibilities for quantum switching, particularly in field-tuning CDW electronics. However, these two phenomena often compete and rely heavily on strong electronic correlations. While carbon nanowire arrays have been experimentally shown to exhibit ferromagnetism above 400 K, our research shows that encapsulating a linear carbon chain (LCC) within zigzag boron nitride nanotubes (BNT) induces a short-range CDW state under a competing effect of ferrimagnetism and magnetic frustrations. However, for this exotic feature to occur, the LCC needs to break the symmetry along the circular plane of the BNT. Then we utilize a Monte Carlo model to identify the optimal length of LCC@BNT to tackle its size effect, while also comparing the stability of chains provided by carbon nanotubes. The shorter LCC@BNT displays a more prominent long-range CDW pattern with a tunneling barrier of 2.3 eV on the Fermi surface, transitioning into an unconventional insulator. Meanwhile, magnetic frustrations disappear, and ferrimagnetism remains stable up to 280 K. Our discovery of ferrimagnetic CDW carbyne insulators, which function without conventional periodic lattice distortion, spin-orbit coupling, or complex d and f hybridization represents a groundbreaking shift in thinking, which demonstrates that such exotic properties are not exclusive to transition metal elements. We anticipate that spin fluctuations in LCC@BNT could enable fine-tuning of the CDW pattern, and applying an electric excitation of 2.3 eV triggers an abrupt insulator-to-conductor transition for quantum switching applications.

## 1. Introduction:

Charge density waves (CDW) and their interactions with ferromagnetism or ferrimagnetism are prominent areas of study in topological condensed matter research, particularly due to the applications of resistive switching and resonant oscillations for topological computing systems [1-5]. However, CDW often competes with non-zero net magnetization (e.g. ferromagnetism, ferrimagnetism), with only a few exceptional cases made up of transition elements documented [1-4]. CDW usually coexists with antiferromagnetism (AFM) [6] such as iron-based superconductors. The technical problem is that the combination of AFM and CDW poses challenges in field-effect applications, as tuning the CDW state with an external field is ineffective in an antiferromagnetic environment due to the net spin being zero [7]. More importantly, a ferromagnetic CDW or ferrimagnetic CDW state can slow down or localize electrons at the Fermi level, resulting in a minimum electronic density of states being zero alternatingly [8]. This leads to a ferromagnetic or ferrimagnetic CDW insulator, which is very rare but could greatly impact spin-filtering tunneling devices [9-12] if the CDW reacts with external fields. These materials can be crucial for advancing quantum information and quantum computations. Apart from this, utilizing CDW modulation to achieve an insulator-to-conductor transition under the control of external fields is a highly desirable function for quantum switching techniques [9-12].

However, many of these desirable exotic properties originate from heavy metals, particularly those dominated by transition metals. The presence of spin-orbit coupling and p-d-f shell hybridization complicates the tuning of these quantum phenomena [13], as any adjustment necessitates a simultaneous readjustment of multiple parameters. This raises questions about whether these complex quantum phenomena can also be realized in systems that rely solely on p-shell electrons, which present a clearer framework for tuning. In this regard, a monoatomic carbon chain, commonly known as carbyne [14-16], represents a promising option for this framework. Although a linearly straight carbon chain is always non-magnetic, 400K ferromagnetism has been experimentally observed in a parallel array of finite-length carbon chains with the help of periodic kink structures [17]. Despite its high Curie temperature, the creation of CDW patterns along these ferromagnetic chains is still an open question.

Worse still, creating bulk carbyne remains a significant challenge despite extensive experimental efforts. Current manufacturing techniques struggle to produce isolated carbon chains longer than 68 atoms, with most isolated chain connected by ~10 atoms [14-16]. A breakthrough was achieved by Lei Shi et al., who used double-walled carbon nanotubes (DWCNT) as nanoreactors to extend carbon chains up to ~6400 atoms [18]. In this setup, the stability of the internal carbyne is influenced by Van der Waals forces, with the CNT wall acting as an effective nanoreactor [18,19]. However, finding the optimal length limit for carbon chains on other substrates or nanoreactors presents a considerable experimental challenge. The inherent linear structure of carbon chains, characterized by alternating strong triple and single covalent bonds, places the carbon atoms in a high-energy state compared to other one-dimensional materials. As the length of the carbon chain increases, the strain and reactivity of the atoms grow exponentially, making the carbon atoms prone to various side reactions. These reactions can quickly disrupt the linear structure and lead to chain breakdown. While the physical and chemical properties of bulk carbon chains can be predicted using repeated units in ab-initio software for valuable insights [14], uncertainty about the actual length presents a significant challenge in correcting for size effects because desired properties may vanish when the carbon chain becomes finite. To carefully consider the size effect, in 2017, C.H. Wong et al. predicted that a parallel chain of carbon could trigger strong ferromagnetism; however, the most probable length and chain-to-chain separation for such configurations remained an open question. The carbon chain array model developed by C.H. Wong et al. [20] can estimate the likely range of chain lengths and lateral separations within array structures, allowing ab initio software to later assess whether the desired properties remain valid at a finite scale. One year later, C.H. Wong et al. validated their carbon chain array model with the correct geometry and confirmed its strong ferromagnetism above 400 K experimentally [17]. In 2023, C.H. Wong et al. developed the carbyne@nanotube model [19], which examines how these nanotube structures influence the stability of the internal carbon chain, yielding a theoretical probable chain length in (6,4)CNT of ~5750 carbon atoms that aligns with experimental findings of ~6400 carbon atoms [18].

To explore magnetic carbon chains within a nanotube structure, carbon nanotubes are not ideal hosts due to the involvement of magnetic transition elements during CNT fabrication [21], which may lead to false magnetic detections of carbon chains. Hence, we are seeking an alternative host, i.e. BNT [22]. However, both theory and experiments suggest that a nanotube host always prefers to keep the internal carbon chain straight that hinders the appearance of magnetism [17-19]. To enable the coexistence of CDW in a magnetic carbon chain and meanwhile allows CDW to respond to external perturbations, we need to address at least seven challenges. First, periodic kink structures are required to induce magnetism in the chain [17]. Since the internal carbon chain within the nanotube always remains straight, we need to come up with another strategy to induce magnetism in the carbon chain without these kink structures. Second, it is essential to ensure that organic magnetism is not antiferromagnetic. Otherwise, it cannot react with external magnetic field for tuning the CDW states. Third, ferromagnetic or ferrimagnetic exchange interactions should be maintained at room temperature, which is often challenging in organic materials. Fourth, the lattice spin must be carefully balanced: sufficiently large to enable couplings with CDW, but not so large that it overwhelms their competing effects [1-4]. Fifth, while CDW is always observed in heavy transition elements [1-4], it is an uphill struggle to create CDW patterns from the light atom (carbon), without relying on strong electronic correlation or spin-orbit coupling. Sixth, we need to determine the optimal length of the internal chain within BNT. If the chain is very short, we have to ensure that its finite length can still maintain the desired properties. Finally, we need to analyze whether the CDW state could be influenced by external perturbation (e.g. magnetic field or electric field) to assess the potential for quantum switching.

## 2. Computational Details

We combine the ab-initio and Monte Carlo methods to simulate the results. Unless otherwise stated, we employ the LDA-PWC functional within the CASTEP Software to relax the LCC@BNT composites [23]. We use a SCF tolerance of 1 x $10^{-5}$ eV and allow up to 9000 SCF cycles to obtain the relaxed geometry of LCC@BNT. The magnetism and CDW are computed by Dmol3 software in BIOVIA Materials Studio. Then the Monte Carlo carbyne@nanotube model [19] is employed to reduce the chain length at a finite scale. If the corrected chain length becomes very short, recalculating the physical properties in the form of a non-periodic supercell is necessary. The ab-initio-calculated atomic coordinates are fed into the Hamiltonian of the carbyne@nanotube model in order to update the atomic coordinates of the internal carbyne at finite temperatures [19].

We revisit the Hamiltonian of the carbyne@nanotube model [19],

$$H = e^{-T/T_{bj}}\left(\sum_{n=1,3,5}^{N}\left|E_{n,j}-E_1\right|e^{-\frac{\ell_n-\ell_{n,j}^{eq}}{0.5\ell_{n,j}^{eq}}}+\sum_{n=2,4,6}^{N}\left|E_{n,j}-E_3\right|e^{-\frac{\ell_n-\ell_{n,j}^{eq}}{0.5\ell_{n,j}^{eq}}}\right)+e^{-T/T_{bj}}\sum_{n=1,2,3}^{N}J_A e^{-\frac{\ell_n-\ell_{n,j}^{eq}}{0.5\ell_{n,j}^{eq}}}\left(\cos\theta+1\right)^2$$
$$-4\varphi\sum_{\phi=0}^{2\pi}\sum_{n=1}^{N}\left[\left(\frac{\sigma}{r}\right)^6-\left(\frac{\sigma}{r}\right)^{12}\right]$$

The adjacent bond distance is $\ell_n$. The lateral distance between LCC and BNT is $r$. The type of covalent bond formed between adjacent carbon atoms is described by stochastic variables j and n, where j represents the bond energies [19] (1 for single bond: $E_1$ = 348 kJ/mol; $\ell_{n,1}^{eq}$ = 154pm, 2 for double bond: $E_2$ = 614 kJ/mol; $\ell_{n,2}^{eq}$ = 134pm , and 3 for triple bond: $E_3$ = 839 kJ/mol; $\ell_{n,3}^{eq}$ = 120pm)[19,20] for the $n^{th}$ carbon atom. For instance, $E_{n,j}$(n = 500, j = 2) refers to the 500th carbon atom (n = 500) forming a double bond (j = 2) with respect to the $499^{th}$ carbon atom (n-1 = 499). The temperature $T_{bj}$ corresponding to bond dissociation = bond energy / Boltzmann constant $k_B$ [19,20]. We initialize the structure of carbyne@nanotube under various chirality numbers ($N_c$,$M_c$). The initial bond type of the internal LCC is j=2. To assess the stability of the carbyne within the BNT environment, we introduce a chain-stability factor [19], defined as $e^{-\frac{\ell_n-\ell_{n,j}^{eq}}{0.5\ell_{n,j}^{eq}}}$. The Van der Waal's VDW term between BNT and LCC along the circular plane is $E_{vdw} = -4\varphi\sum_{\phi=0}^{2\pi}\sum_{n=1}^{N}\left[\left(\frac{\sigma}{r}\right)^6-\left(\frac{\sigma}{r}\right)^{12}\right]$. The $\sigma$ and $\varphi$ are the VDW constants [19,20] which are interpreted from isothermal compressibility and the sample length [19,20]. Based on the ab-initio data of a kink structuring carbyne chain, the angular energy term is calibrated by fitting $\left(\cos\theta+1\right)^2$ function, with the $J_A$ of ~600 kJ/mol [19,20] and the pivot angle $\theta$ along the carbon chain.

To simulate the dynamic behavior of the carbyne chain within the nanotube environment, we employ the Monte Carlo approach. During the iterative process, the atomic coordinates and the types of covalent bonds in the carbon chain are amended at finite temperatures. In each Monte Carlo step (MCS), the simulation follows these steps [19]:

- Randomly select an atom in the carbon chain.
- Calculate the initial Hamiltonian of the system.
- Propose a trial range of spatial fluctuations and a trial type of covalent bond (Figure S1 in supplementary materials) for the selected atom.
- Estimate the trial Hamiltonian based on the proposed changes.
- If the trial Hamiltonian is less positive (or more negative) than the initial Hamiltonian, the trial states are accepted. Otherwise, the system reverts to the previous states.
- The trial range of atomic fluctuations is also controlled by the Hooke's factor (supplementary materials).

Thermal excitation is another opportunity to accept or reject the trial states in parallel, which is governed by the comparison between a random number with the Boltzmann factor [19,20]. The Monte Carlo MCS process continues until the energy vs MCS becomes flattened, typically after approximately 250,000 steps based on our previous result [19]. The data is then averaged over the last 20000 equilibrium data points. Furthermore, we analyze the optimal (or cut-off) length of the internal chains in BNT versus CNT. This approach allows us to compare the stability and length of LCC within distinct nanotube structures. After correcting the size effect by the carbyne@nanotube model, we estimate the exotic properties of the internal carbon chain again.

## 3. Results and Discussion

### 3.1. The long LCC@BNT as a ferrimagnetically frustrated CDW conductor

Despite an isolated straight carbon chain is always non-magnetic [17], our search for magnetism in a straight carbon chain begins with introducing the proximity interactions from BNT. The carbon chain is inserted in zigzag nanotubes ($N_c$,$M_c$) with the diameters between ~0.4nm and ~1.7nm, where $N_c$ is from 5 to 21 and $M_c$ is 0. For $N_c < 11$, the internal carbon chain is symmetric along the circular plane of the BNT. However, when $N_c \geq 11$, the internal carbon chain automatically breaks this symmetry and displaces towards the BNT surface in the inset of Figure 1a. If the tube diameter is small, the internal chain benefits from a uniform lateral Van der Waals force along the angular plane. However, when the tube diameter exceeds ~0.9 nm (or $N_c \geq$ ~11), the internal chain loses the uniform Van der Waals protection from the nanotubes. We observe that the internal chain tends to partially regain Van der Waals protection by displacing the internal carbon chain (inset of Figure 1a) radially. Without this radial displacement, the carbon chain risks losing all its Van der Waals protection for maintaining stability within the nanotube structure. By adjusting its geometric arrangement radially, the carbon chain can still interact with the surrounding nanotube walls, allowing it to harness some of the lateral Van der Waals forces. This partial Van der Waals interaction helps to provide a degree of stability and support, even in the absence of the full protective effect initially present at smaller diameters.

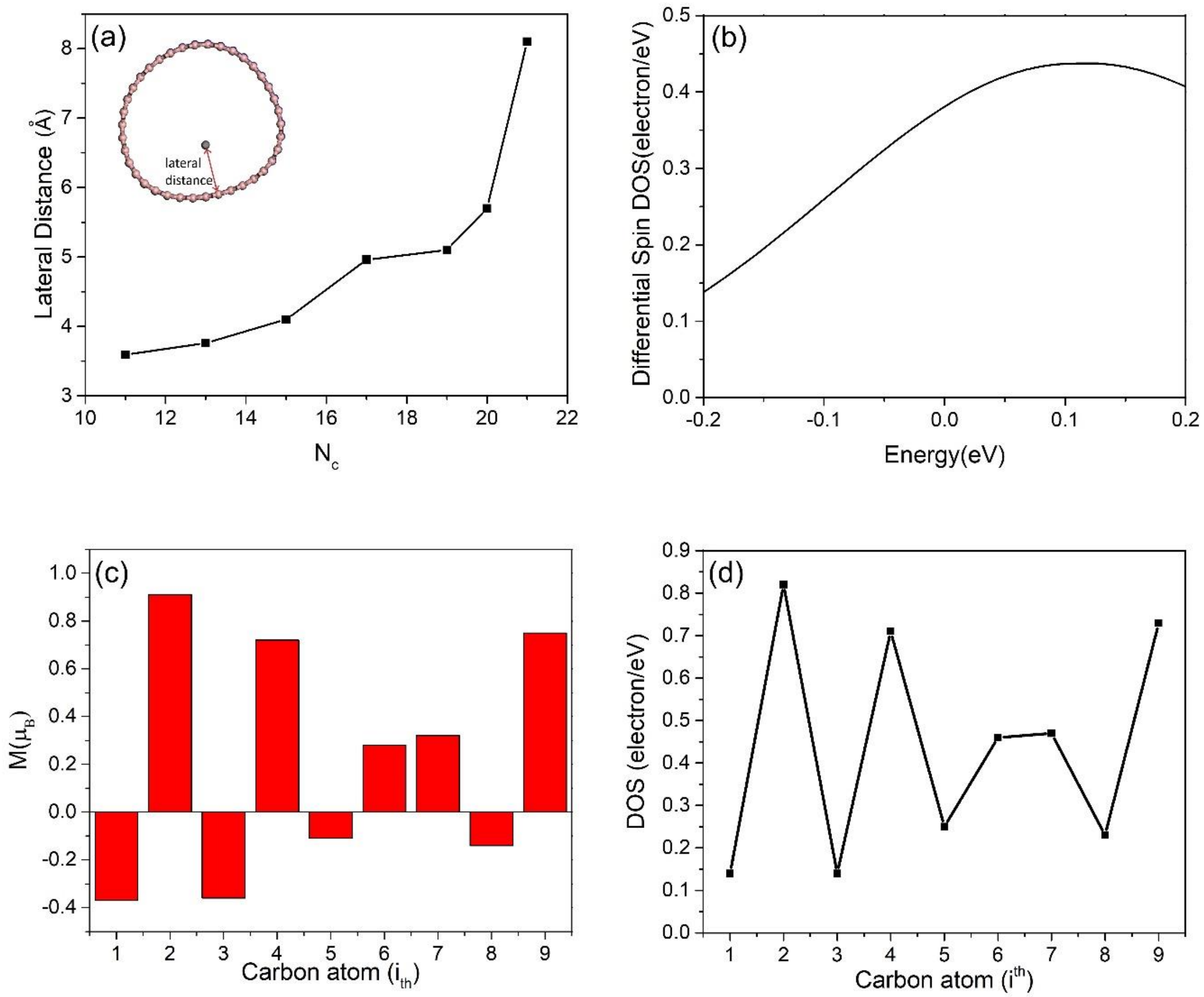

Figure 1: (a) After geometric optimization, the nearest lateral distance between the infinitely long carbon chain and the zigzag BNT is depicted. The inset shows a cross-section of an infinitely long carbon chain enclosed by a (19,0) boron-nitride nanotube, where the chain axis is along the out-of-plane direction. The 9-atom-long carbon chain enclosed by BNT forms a repeated unit. The grey ball is carbon, the pink balls are the nanotube. (b) The differential spin DOS per carbon atom in LCC@(19,0)BNT. This is 100% p-orbital magnetism. The Fermi level ($E_F$) is shifted to 0eV for convenience. (c) The distribution of lattice spin exhibits signs of magnetic frustrations. While the 7$^{th}$ and 9$^{th}$ carbon atoms are expected to have a spin-down state, the 8$^{th}$ is anticipated to have a spin-up state. If the 9$^{th}$ carbon atom is spin down, the 1$^{st}$ atom in the repeated unit cannot also spin down when ferrimagnetism prevails. Consequently, magnetic frustrations introduce disorder in the spin distribution within the repeated unit. (d) A long-range CDW pattern is disrupted by magnetic frustrations, as indicated by the uncertainty in lattice spin in the corresponding regions.

The pivot angles between the carbon chains range from 179.80 to 179.96 degrees for the tube diameters analyzed in Figure 1a, suggesting that the chains can be regarded as straight [17]. Therefore, we cannot apply the periodic kink-structure strategy [17] to induce magnetism in the carbon chains, which typically requires a threshold pivot angle sharper than 170 degrees [17].

Our simulations indicate that only the (17,0)BNT and (19,0)BNT exhibit magnetism. The average magnitude of magnetic moment of the carbon chain in (17,0) BNT and (19,0) BNT are both ~0.6 $\mu_B$, which is at least three times larger than the average magnetic moment of the kink-structured carbon chains [17]. Moreover, the local lattice spin of the internal carbon chain in (19,0) BNT is slightly greater, measuring ~0.9$\mu_B$, compared to ~0.7 $\mu_B$ in (17,0) BNT, but no magnetic properties are observed in the B and N atoms. Consequently, we focus solely on the average spin DOS of the C atoms in Figure 1b, which illustrates their magnetic nature. Using a 9-atom-long carbon chain enclosed by BNT forms a repeated unit, rather than using a basic unit of 3-atom-long carbon chain, because it more clearly draws the oscillations along the chain axis. If a 2-atom-long (or 4-atom-long) carbon chain is used per repeated unit, the carbon atoms are too far apart (or too close together) to form effective trial covalent bonds. Consequently, those configurations fail to relax the system to its ground state in our ab initio simulations. Since the local magnetism of the carbon atoms exhibits a stronger magnetic moment inside (19,0)BNT, we analyze the distribution of magnetic moments of the chain within (19,0)BNT in Figure 1c, where we observe the two spin directions of carbon atoms compete, leading to the emergence of magnetic frustration, akin to Kagome lattice [24].

We observe a short-range CDW pattern in the carbon chain within the (19,0)BNT between the first and third carbon atoms in Figure 1d. The CDW structure should have restored a DOS($E_F$) of 0.83 electron/eV and 0.14 electrons/eV for the $4^{th}$ and $5^{th}$ carbon atom, respectively. However, magnetic frustration between the sixth and seventh carbon atoms disrupts the long-range establishment of the CDW. It is because the differential electron (charge) density at the Fermi surface triggers an equivalent effect of local magnetism based on Maxwell's equations, which is further influenced by the lattice spin distribution. As a result, the DOS($E_F$) in the vicinity of the magnetically frustrated region is affected, where the "bridge" is the local magnetism. This also explains why the DOS($E_F$) at the $5^{th}$ carbon and $8^{th}$ differs from that of the first and third carbon atoms. This CDW, despite its short range, distinguishes our findings from a conventional CDW pattern because the carbon chain is in cumulative phase (same bond length), thereby ruling out the possibility of being a Peirels-distorted insulator [25]. The long carbon chain in zigzag BNT prefers a cumulative phase over a polyyne phase because the creation of alternating single and triple bonds leads to conflicts in bond type and geometry under a translational symmetry. This is similar to how spin-up and spin-down configurations conflict with geometry. When these bond and location conflicts arise, a significant number of lone pair electrons are generated from the internal chain, undermining its stability.

The non-zero magnetic moment of carbon, despite its linear configuration, indicates that the proximity effect from the BNT is influential. This is particularly noteworthy since we do not observe a similar effect in CNT hosts of the same diameter. The nearest lateral distance illustrated in Figure 1a indicates that the proximity interaction between the BNT and the carbon chain is maximized at a radial separation of ~0.5 nm for both (17,0)BNT and (19,0)BNT. Both the straight carbon chain and the isolated boron-nitride nanotube are originally non-magnetic in their separate systems. To explore the proximity effect between these non-magnetic materials and how one of them (carbon chain) can become magnetic, we compare the cases of "LCC@(19,0)BNT vs LCC@(19,0)CNT", and "LCC@(17,0)BNT vs LCC@(17,0)CNT". To ensure that the atomic coordinates remain consistent in these composites, we replace the B and N atoms with C atoms without conducting another round of geometric optimization. This approach ensures that the proximity effect can be compared under the same atomic coordinates. In these comparative studies, the LCC@(19,0)CNT and LCC@(17,0)CNT can be considered as non-magnetic. In other words, the intrinsic electric field established by the Group III (boron) and Group V (nitrogen) elements could be responsible for creating a magnetic carbon chain, even in the absence of a periodic kink structure. In the previous study, we observed that the periodic E-field from dopants could also trigger magnetism along the carbon chain [26]. Although there are no dopants in the internal carbon chain, the B and N atoms have built a periodic E-field, which resembles the effects of on-site potential modulation [17]. The periodic kink-structured 400K ferromagnetic carbon chains are laterally spaced by 0.5 nm in the experiment [17], which is consistent with the lateral distance observed in the two magnetic LCC@BNT composites (Figure 1a).

### 3.2. Length limitation in LCC@BNT

Before we predict the finite size effects of magnetism and charge density wave, we have to determine how long the carbon chain can exist within the BNT. The carbyne@nanotube model [19] can be considered as a reasonable tool to assess the stability factor of the internal carbon chain. By replacing CNT with BNT in this model, we investigate the stability of a finite-length LCC as a function of the BNT diameter. The carbon chain exhibits the strongest mechanical strength (i.e. very strong covalent bond), if not the absolute strongest [14-15]. However, its magnetic moment is only a fraction of the Bohr magneton, indicating that the magnetic energy is significantly weaker than the covalent bond between the carbon atoms. As a result, its weak magnetic interaction is reasonably not included in the Hamiltonian. While the longest LCC achieved to date has been through encapsulation in a (6,4)CNT [18], our analysis begins by comparing the optimal LCC length within (6,4)CNT versus (6,4)BNT to observe whether there are significantly distinct levels of protection and subsequently interprets the optimal LCC length within (19,0)BNT. Back to the carbyne@nanotube model of LCC@(6,4)CNT, a more rapid downturn in the chain stability factor of ~0.7 is observed above ~5750 carbon atoms [19], which almost matches the experimental cut-off LCC length of ~6400 atoms [18]. Applying the same criteria (chain stability factor ~ 0.7) to judge the cut-off (or called optimal or most probable) LCC length inside (6,4)BNT at 300K, the estimated value drops to only ~900-atom-long in Figure 2a. In other words, the use of BNT is unlikely to suit long carbon chain production under the same tube diameter and chirality. The stability factor of ~0.7 is equivalent to an average bond length of 1.39Å, close to the average single and triple carbon-carbon bond distance of (1.21+1.54)/2 = 1.37Å. The lower stability factor of ~0.55 corresponds to an average bond length of ~1.50Å, still less than the maximum permittable C-C length ~ 1.73Å [27]. The finite polyyne chain with N = 900 carbon atoms exhibits increasing stability as the temperature is lowered, as shown in Figure 2b. This phenomenon can be explained by the reduction in thermal fluctuations at low temperatures. When the temperature is lowered, the thermal energy available to the atoms decreases. As a result, the carbon atoms within the polyyne chain vibrate less vigorously. This reduced atomic motion allows the carbon atoms to favor and occupy their ground state more steadily. At low temperatures, the minimization of thermal fluctuations leads to a higher degree of stability for the polyyne chain. The carbon atoms are less likely to be displaced from their optimal positions, and the overall structure of the chain becomes more stable.

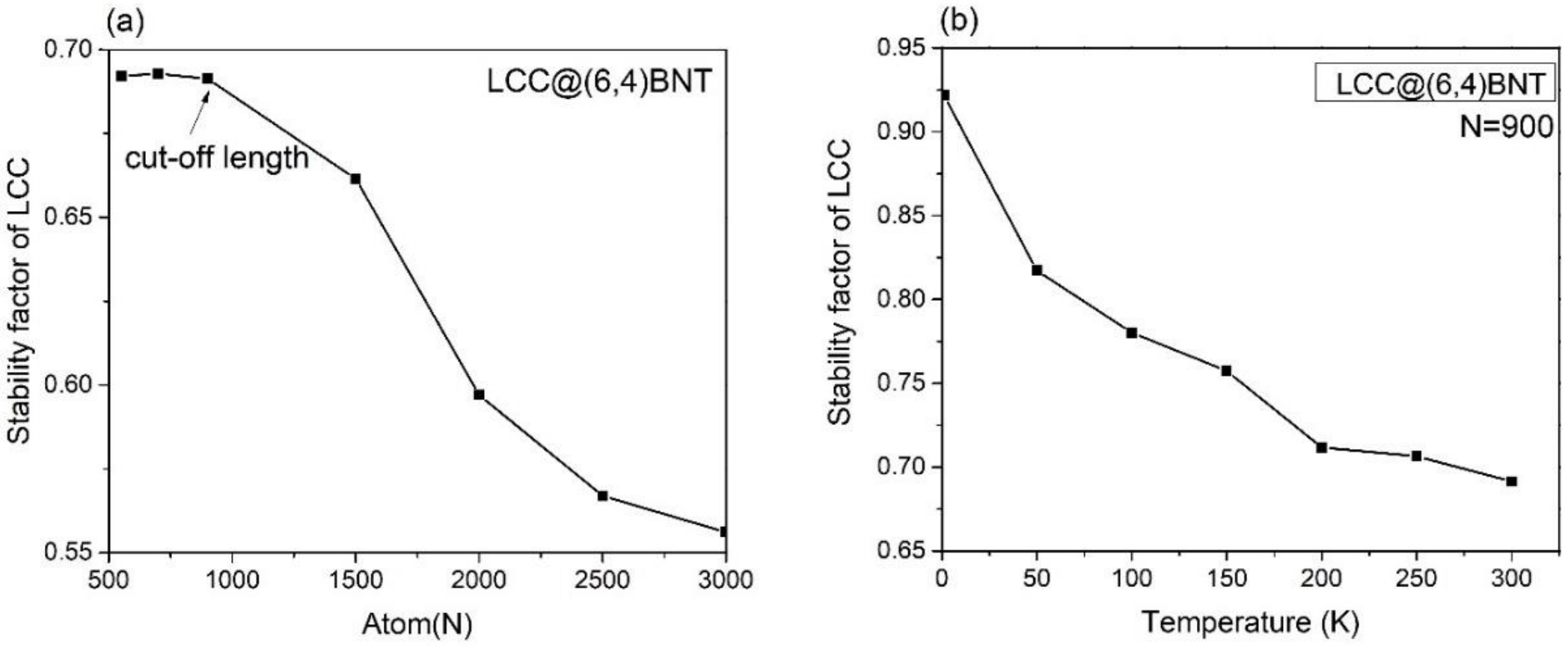

**Figure 2.** (a) The length dependence of the chain stability of LCC@(6,4)BNT at 300K; (b) Reduction in the stability factor is observed in LCC@(6,4)BNT.

According to Figure 3a, the formation of the polyyne phase ($[C-C\equiv]$) is energetically favorable [18] in 900-atom-long LCC inside (6,4)BNT, even up to room temperatures. The percentage of the polyyne phase remains above 95% even though we observe that the polyyne phase starts to pale under thermal excitations. Correspondingly, as shown in Figure 3b, the formation of the cumulene phase (C=C bonds) is tiny. Our Monte Carlo simulator also detects rare signals, less than 0.1%, for the formation of $[C-C=]$ or $[-C-]$ phase. We further study and examine the relationship between the cut-off LCC length and the chirality numbers of BNT in Figure 3c. The results indicate that as the BNT radius increases from approximately 0.25 nm to 0.5 nm, the cut-off LCC length shows a relatively linear decrease. The curve then takes on an exponential-like shape, approaching around 10 atoms for (19,0)BNT. The abrupt change in the slope at $N_c$ = 12 is due to the formation of an asymmetric shape of BNT (see the inset of Figure 1a). No data is computed for BNT structures thinner than (5,0), as it may trigger the formation of covalent bonds between the carbon chain and the BNT surface which would violate the underlying assumption of the Hamiltonian, where only van der Waals interactions exist between the carbon chain and the BNT.

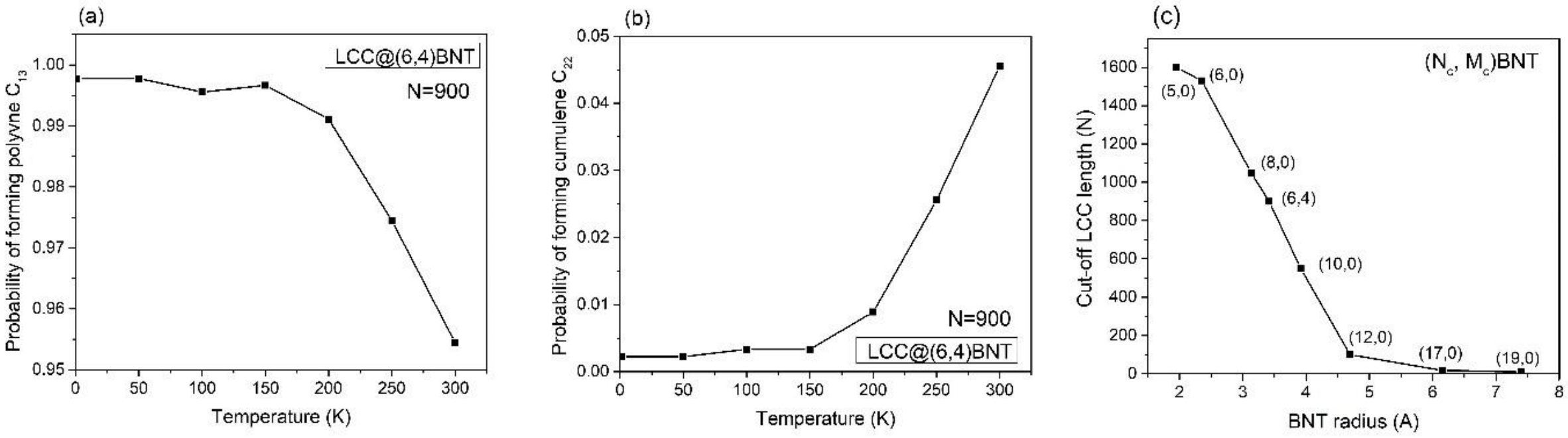

**Figure 3. (**a) The probability of the polyyne phase (alternating single and triple bonds) slightly drops from around 0 K up to 150 K. However, above 150 K, the probability of the polyyne phase drops more rapidly. (b) The probability of the cumulene phase (consecutive double bonds) increases very slowly from around 0 K up to 150 K. But above 150 K, the formation of the cumulene phase becomes much faster. (c) The cut-off length of LCC encapsulated by ($N_c$,$M_c$)BNT at 300K.

We further analyze the cut-off length of the LCC within CNT versus BNT in the presence of a Group VII dopant. Doping with a Group VII element into nanotube is anticipated to shorten the LCC length due to the increased number of lone pair electrons on the nanotube, which leads to enhanced electrostatic repulsion. Therefore, the test specimen, fluorine, is selected as the dopant for the nanotube. We arbitrarily select $N_c$ =10 and compare the cut-off LCC lengths in (10,0) CNT versus (10,0) BNT. In the case of (10,0)BNT, approximately 1% fluorine doping dramatically reduces the LCC length from 600 to less than 100 atoms. In contrast, the LCC length in (10,0) CNT only decreases from about 1000 to 800 with the same level of fluorine doping. These results suggest that impurity control is more critical for growing LCC within BNT compared to CNT hosts.

The choice of setting the maximum Monte Carlo steps (MCS) to 250,000 has been justified [19], as this is large enough to allow the composite system to reach equilibrium. This is supported by the evidence presented in Figure 5 of reference 19, which shows that the energy versus MCS plot for a 15,000-atom-long carbyne inside a carbon nanotube (CNT) has reached equilibrium for MCS values exceeding ~180,000 [19]. In the current study, the longest internal carbyne is only up to ~1000 atoms. Since the required MCS duration is proportional to the chain length of carbon, the chosen MCS duration of 250,000 steps is suitable for this investigation.

### 3.3. The short LCC@BNT as a ferrimagnetic CDW insulator

Based on the LCC@(19,0)BNT model, we anticipate that the internal carbon chain is connected by approximately 10 atoms. Hence, we constructed a non-periodic supercell with 9 carbon atoms surrounded by (19,0) BNT. After geometric optimization of the finite supercell, the proximity effect resulted in the nearest lateral distance being ~0.55 nm, which falls within the expected range for magnetism, where it breaks the symmetry along the circular plane of BNT again. However, no magnetism is observed in this internal carbon chain surprisingly. We suspect that the disappearance of the internal finite-length LCC is due to the formation of free radical electrons at the opposite ends of the chain (4 free radical electrons are detected). Therefore, to activate the re-emergence of magnetism, we employ hydrogen termination, which is a common method for terminating chain growth [28]. Geometric optimization of finite-length $H\text{-}C_9\text{-}H$ within (19,0)BNT is conducted again under the same ab-initio setup, where the supercell is drawn in the inset of Figure 4a. After hydrogen termination, the carbon atoms reappear magnetism in the supercell structure. Ferrimagnetic state is observed in the finite carbon chain in (19,0)BNT without magnetic frustrations, as depicted in Figure 4a and Figure 4b. The finite size effect causes a reduction in the highest local magnetism from 0.9 $\mu_B$ to 0.6 $\mu_B$, where the strong exchange interaction is ~24meV (or ~280 K). When compared to the infinitely long case, Figure 4c shows a more pronounced long-range CDW pattern along the entire carbon chain, with a complete cut-off of DOS at the even-numbered carbon sites, which is a sign of unconventional insulators [9,11]. To enable the electron tunnel between odd-numbered carbon sites on the Fermi surface, the tunneling barrier of 2.3 eV needs to be overcome as shown in Figure 4d.

The $H\text{-}C_9\text{-}H$ within (19,0)BNT contains over ~300 atoms in the supercell which raises a high computational cost. Fortunately, in 2018, we analyzed the magnetic states of carbon chains and discovered that using either the LDA or GGA level provides a very accurate prediction of experimental ferromagnetism above room temperatures [17]. Similarly, this magnetism focuses exclusively on carbon chain again, thereby avoiding the complexities associated with transition metals. Additionally, no spin-orbit coupling is observed in the internal carbon chains. Hence, our selected DFT functional is a sensible choice [23], as it highlights that the coexistence of magnetism and CDW effects in the carbon chain does not involve complex electron-electron interactions. Dmol3 and CASTEP are accurate ab-initio software for nanowires [28-32], where their accuracies have been justified [32]. The average magnitude in magnetic moment of LCC@(19,0)BNT computed using LDA-PWC in Dmol3 software is approximately 0.6 μB, while GGA-PW91 or GGA-PBE in CASTEP software yields around 0.5 μB. The cut-off LCC length calculated using GGA-PW91, GGA-PBE, LDA-PWC, and LDA-VWN methods is consistent, showing only a 1-2% variation in the computed length only. The CDW patterns obtained using GGA-PW91, GGA-PBE, LDA-PWC, and LDA-VWN are also consistent, indicating that the computed results are not highly sensitive to the choice of DFT functional or software.

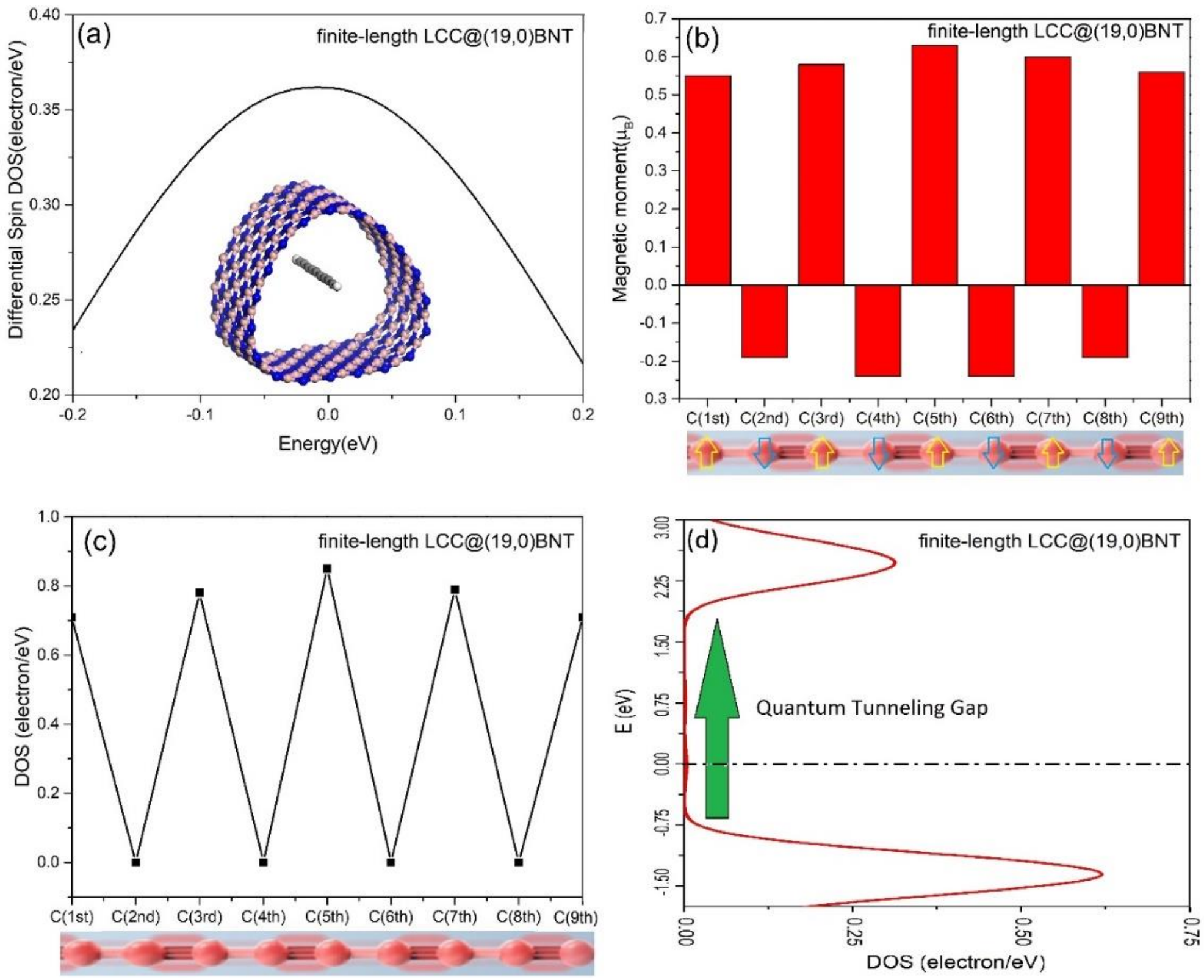

**Figure 4**: (a) The differential spin DOS of the internal H-$C_9$-H chain per atom. This is 100% p-orbital magnetism. The Fermi level is shifted to the Fermi level for convenience. The inset shows the non-periodic supercell structure, where the grey balls represent carbon, the blue balls represent nitrogen, the pink ball represents boron, and the white balls represent hydrogen; (b) A clear ferrimagnetic distribution of lattice spin in the finite-length carbon chain; (c) A long-range CDW present in the entire chain in form of an unconventional insulator; (d) The DOS plot of an even-numbered carbon atom indicates a deficiency of Fermi electrons. To enable the electron tunnel between odd-numbered carbon sites on the Fermi surface, the tunneling gap of ~2.3 eV needs to be overcome.

### 3.4. Field Tuning of Insulator-to-Conductor Transition and Further Scientific Insights

Looking ahead, we foresee a future in which organic materials could play a role in the development of next-generation field-tuning CDW devices, leading to novel spin transport [1-12]. Comparing the distinct CDW patterns in Figure 1 and Figure 4, we observe that magnetic fluctuations can disrupt a long-range CDW pattern, where ordered spins promote ferrimagnetism and long-range CDW on a finite scale. The sharp gradients in DOS($E_F$) within the CDW structure generate local magnetism equivalently, in line with Maxwell's equations. Consequently, an exotic coupling arises between this local magnetism and the underlying magnetic background. Thus, applying a magnetic field perpendicular to the easy axis of ferrimagnetic alignments induces spin fluctuations that anticipate CDW modulation, modifying the electronic properties.

To produce an abrupt insulator-to-conductor transition, should a magnetic field or an electric field be applied? Based on Figure 1, applying an external magnetic field could help adjust the CDW pattern associated with electronic properties by triggering subtle spin fluctuations, but it may not suit to produce a rapid insulator-to-conductor transition. Although Figure 1 involves an infinite long LCC@BNT, the underlying science regarding to the subtle spin fluctuations should align with the finite scale. Conversely, applying a constant electric excitation of 2.3 V along the chain axis is a more promising approach for overcoming the tunneling barrier and achieving a rapid insulator-to-conductor transition, as we see it in the finite case (Figure 4). Using a constant electric excitation to trigger the insulator-to-conductor transition is more superior than using external magnetic field because the external magnetic field may introduce magnetic noise within the ferrimagnetic carbon chain.

This research leads to more exotic findings in organic science. Contrary to our initial expectation that a linearly straight carbon chain would not exhibit magnetism [17], we discovered that interactions resulting from the broken symmetry in non-magnetic BNT create magnetic proximity to the carbon chain. This is highly unusual, but it suggests that, in the field of organic science, there may be multiple pathways to trigger magnetism in organic substances. The periodic kink structures in a short carbon chain arranged in parallel has shown ferromagnetism experimentally. In contrast, a short carbon chain enclosed by BNT could generate ferrimagnetism, while increasing the length of the internal carbon chain within BNT emerges magnetic frustration [24]. By replacing the substrate and scaling the carbon chain, we could alter the type of magnetism in carbon chain, highlighting its vast potential applications in spintronics. In addition, we have found that CDW and ferrimagnetism are compatible in this sample, whereas ferromagnetism and CDW are not compatible here. This indicates that the co-occurrence of ferromagnetism and CDW presents a much tougher challenge comparably [1-4]. The emergence of CDW patterns in full organic material is particularly noteworthy because these unusual properties are always found in transition elements only. Our findings [37] demonstrated that carbyne has the potential to enter the realm of topological science when the substrate is carefully selected. Notably, a Dirac gap at M point is observed in Ru-metalated carbyne under a ruthenium substrate [37]. Interestingly, when a different substrate (BNT) is chosen, carbyne surprisingly exhibits a charge density wave (CDW) state. This creates a new way of thinking that much more exotic science in organic materials could be uncovered through the exploration of other carbyne-substrate combinations.

Conventional CDW patterns typically require periodic lattice distortion, which leads to the formation of an ordered quantum fluid of electrons at the Fermi level [4]. However, the cumulative phase does not exhibit periodic lattice distortion. Hence, our discovery of the CDW pattern without periodic lattice distortion in section 3.1 marks a significant departure from conventional CDW patterns. Shortening the internal carbon chain to a finite length ultimately transforms the system into polyyne, where periodic lattice distortion occurs. The CDW pattern in Figure 4c appears as a superposition of both conventional and unconventional cases. Upon examining the DOS values, the unconventional case dominates. The combined effect of both conventional and unconventional cases results in a more pronounced CDW effect, leading to an alternating complete vanishing of DOS($E_F$). In Figure 4c, the presence of zero density of states at even-numbered sites supports the identification of ferrimagnetic insulators [10, 11]. These materials could act as spin-filtering tunneling barriers for tunneling magnetoresistance [12]. While ferrimagnetic insulators could allow the transport of spin momentum without charge, this unique property has attracted considerable interest in dissipation-less electronic and spintronic devices, solid-state quantum computing, and magnetic tunneling junctions [1-5].

We acknowledge that creating a carbon chain inside single-wall BNT is a challenging task. However, this does not imply that it is without hope. In an experiment conducted by Ryo Nakanishi et al [38], the successful fabrication of single-wall BNT with a very small radius of 0.35 nm inside a CNT represents a crucial first step toward achieving our goal. Scientific advancements often come with difficulties, but the experimental support [18, 38] they provide keeps hope alive. We believe that with continued effort, the creation of a carbon chain inside single-wall BNT should be possible. Each step forward in our research not only enriches organic science but also inspires optimism for future breakthroughs. The challenges we face are substantial, but they may not be insurmountable, as foundational frameworks (experimental fabrications of LCC@CNT and BNT@CNT) are already in place [18,38]. We remain confident that our efforts will lead to new scientific horizons.

**Conclusion:**

Our study offers valuable insights to reconsider the conventional belief that organic materials cannot be promising candidates for emerging exotic charge density waves (CDW). We demonstrate that the broken symmetry of a long carbon chain within boron nitride nanotubes can induce a short-range CDW alongside a ferrimagnetically frustrated state. After addressing the size effects of the carbon chain, we find that these exotic properties persist in finite-length carbon chains, revealing a transition to a pure ferrimagnetism with a long-range CDW state, characterized by a strong exchange interaction of ~24meV. The coexistence of ferrimagnetism and CDW within the LCC@BNT configuration anticipate that an orthogonal magnetic field could tune the CDW states through spin fluctuations, while applying an electric field favors an abrupt insulator-to-conductor transition. This underscores the vast potential for quantum switching and spin computing applications.

**Acknowledgement:**

The authors thank the Department of Industrial and Systems Engineering for providing computational support.

**Conflict of Interests:**

The authors declare no conflict of interests

**Data Availability Statement:**

Data availability is possible upon reasonable request.

**Author Contributions:**

Conceptualization: C.H.W; Methodology: C.H.W; Computation: C.H.W; Validation: C.H.W; Formal analysis: C.H.W, L.Y.F.L, X.H; Supervision: C.H.W; Writing manuscript: C.H.W; Editing manuscript: C.H.W, Z.L.G, C.Y.T, Y.H.T; Software: C.H.W, C.Y.T; Data curation: C.H.W. Z.L.G; Visualization, C.H.W, K.C.L, C.P.C, W.Y.C; Resource: C.H.W, C.Y.T, Y.H.T, L.Y.F.L, X.H.

# Supplementary Materials

Guide to propose the trial atomic fluctuations of the internal carbyne at each MCS

- The trial fluctuations (δx, δy, δz) of the randomly selected carbon atom along x, y and z axis:

  $$\delta x = \pm v < \delta t > R_c$$

  $$\delta y = \delta z \sim \frac{k_B T}{< E_1 + E_2 + E_3 >} \delta x$$

  The free particle velocity is $v = \sqrt{\frac{k_B T}{M}}$.

  The average scattering time $< \delta t >$ ~ $10^{-13}$s

  M: The mass of a carbon atom.

  $R_c$: Random number within 0 and 1

  $k_B$: Boltzmann constant

  $< E_1 + E_2 + E_3 >$: The average energy of single, double and triple bond.

- The stable carbyne chain inside the host should have a shorter mean free path (MFP) and a narrower range of spatial fluctuation compared to an unstable isolated carbyne chain. Hence, the Hooke's factor $f(Hooke)$ is created to refine the trial atomic fluctuation.

  $$f(Hooke) = \frac{K_{break}\left(\ell_{\max} - < \ell_{eq} >\right)}{K_{carbyne@host}\left(\ell_{carbyne@host}\big|_{0K} - < \ell_{eq} >\right)}$$

  The refined trial dynamics (δx', δy', δz') = ( $\delta x \cdot f(Hooke), \delta y \cdot f(Hooke)$, $\delta z \cdot f(Hooke)$ ) monitoring from a chain-breaking state to a crystalline form.

  Chain-breakage: the maximum $C-C$ length = $\ell_{\max}$ ~ 1.73 Å associated with the spring constant $K_{break}$.

  The ground state of an isolated LCC: bond length = $< \ell_{eq} > \sim$ 1.34Å

  The average bond length of the internal LCC in the host at 0K = $\ell_{carbyne@host}\big|_{0K}$ associated with the spring constant $K_{carbyne@host}$.

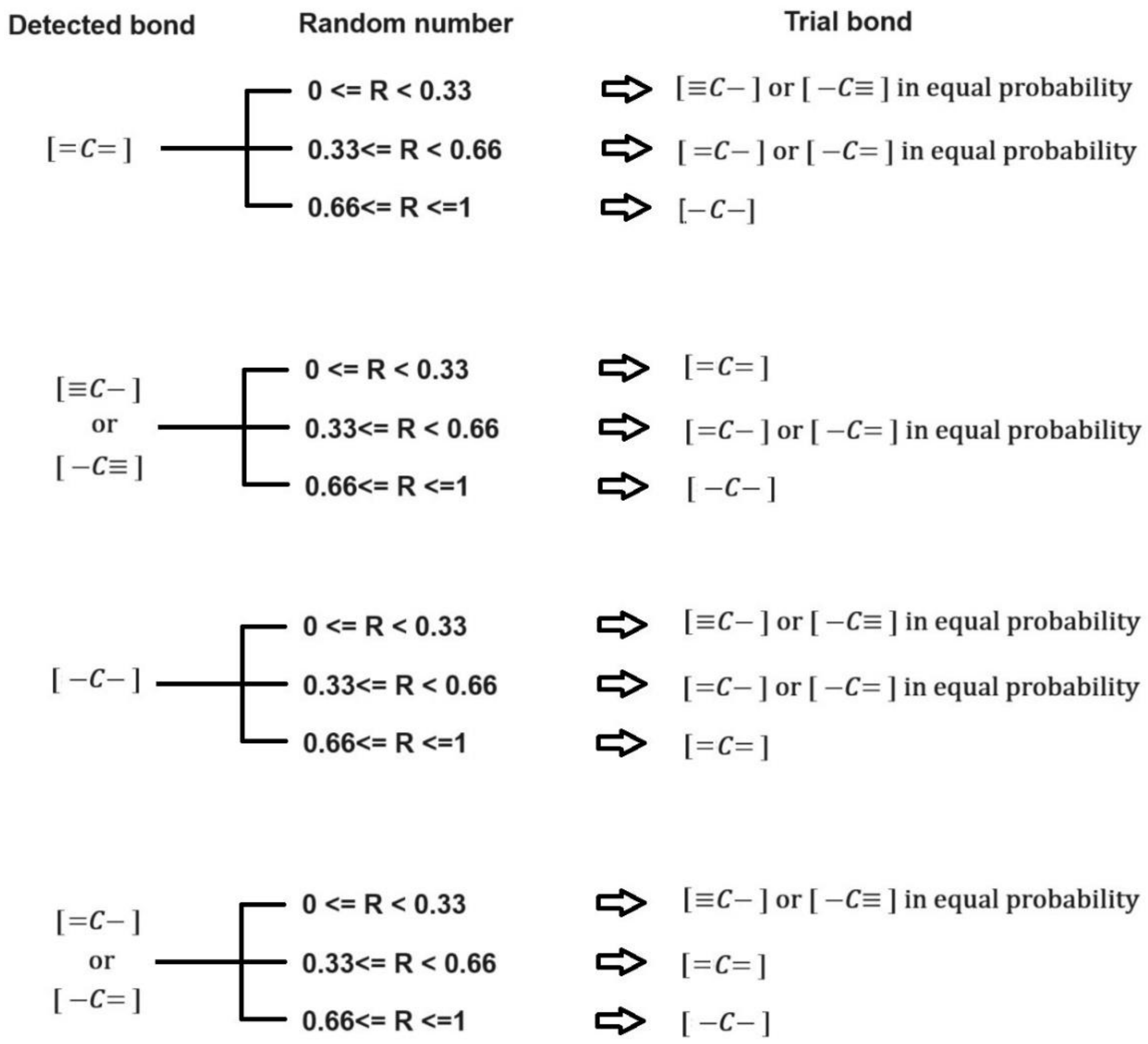

Figure S1: Guide to propose the trial type of covalent bond of the internal carbyne at each MCS [14]